\documentstyle[preprint,eqsecnum,epsf,aps,floats,tighten]{revtex}
\input epsf

\newcommand{\XX}{\mbox{$\, \times \,$}}
\newcommand{\pbarp}{\mbox{$\bar{p}p$}}

\newcommand{\cmehi}{\mbox{${\sqrt{s}=1800}$\,GeV}}
\newcommand{\cmelo}{\mbox{$\sqrt{s}=630$\,GeV}}
\newcommand{\eboth}{\mbox{$\sqrt{s}=630$ and $1800$\,GeV}}
\newcommand{\rs}{\mbox{$\sqrt{s}$}}
\newcommand{\ncal}{\mbox{$n_{{\rm CAL}}$}}
\newcommand{\nl}{\mbox{$n_{{\rm L\O}}$}}

\begin{document}
\title{
Hard Single Diffraction in $\bbox{\bar{p}p}$ Collisions at 
$\bbox{\sqrt{s}}$ = 630 and 1800 GeV
}

%
\author{                                                                      
B.~Abbott,$^{47}$                                                             
M.~Abolins,$^{44}$                                                            
V.~Abramov,$^{19}$                                                            
B.S.~Acharya,$^{13}$                                                          
D.L.~Adams,$^{54}$                                                            
M.~Adams,$^{30}$                                                              
S.~Ahn,$^{29}$                                                                
V.~Akimov,$^{17}$                                                             
G.A.~Alves,$^{2}$                                                             
N.~Amos,$^{43}$
% ...
\\                                                                            
\vskip 0.30cm                                                                 
\centerline{(D\O\ Collaboration)}                                             
\vskip 0.30cm                                                                 
}                                                                             
\address{                                                                     
\centerline{$^{1}$Universidad de Buenos Aires, Buenos Aires, Argentina}       
\centerline{$^{2}$LAFEX, Centro Brasileiro de Pesquisas F{\'\i}sicas,         
                  Rio de Janeiro, Brazil}                                     
\centerline{$^{3}$Universidade do Estado do Rio de Janeiro,                   
                  Rio de Janeiro, Brazil}                                     
\centerline{$^{4}$Institute of High Energy Physics, Beijing,                  
                  People's Republic of China}                                 
\centerline{$^{5}$Universidad de los Andes, Bogot\'{a}, Colombia}             
\centerline{$^{6}$Universidad San Francisco de Quito, Quito, Ecuador}         
\centerline{$^{7}$Institut des Sciences Nucl\'eaires, IN2P3-CNRS,             
                  Universite de Grenoble 1, Grenoble, France}                 
\centerline{$^{8}$Centre de Physique des Particules de Marseille,             
                  IN2P3-CNRS, Marseille, France}                              
\centerline{$^{9}$LPNHE, Universit\'es Paris VI and VII, IN2P3-CNRS,          
                  Paris, France}                                              
\centerline{$^{10}$DAPNIA/Service de Physique des Particules, CEA, Saclay,    
                  France}                                                     
\centerline{$^{11}$Panjab University, Chandigarh, India}                      
\centerline{$^{12}$Delhi University, Delhi, India}                            
\centerline{$^{13}$Tata Institute of Fundamental Research, Mumbai, India}     
\centerline{$^{14}$Seoul National University, Seoul, Korea}                   
\centerline{$^{15}$CINVESTAV, Mexico City, Mexico}                            
\centerline{$^{16}$Institute of Nuclear Physics, Krak\'ow, Poland}            
\centerline{$^{17}$Institute for Theoretical and Experimental Physics,        
                   Moscow, Russia}                                            
\centerline{$^{18}$Moscow State University, Moscow, Russia}                   
\centerline{$^{19}$Institute for High Energy Physics, Protvino, Russia}       
\centerline{$^{20}$Lancaster University, Lancaster, United Kingdom}           
\centerline{$^{21}$University of Arizona, Tucson, Arizona 85721}              
\centerline{$^{22}$Lawrence Berkeley National Laboratory and University of    
                   California, Berkeley, California 94720}                    
\centerline{$^{23}$University of California, Davis, California 95616}         
\centerline{$^{24}$California State University, Fresno, California 93740}     
\centerline{$^{25}$University of California, Irvine, California 92697}        
\centerline{$^{26}$University of California, Riverside, California 92521}     
\centerline{$^{27}$Florida State University, Tallahassee, Florida 32306}      
\centerline{$^{28}$University of Hawaii, Honolulu, Hawaii 96822}              
\centerline{$^{29}$Fermi National Accelerator Laboratory, Batavia,            
                   Illinois 60510}                                            
\centerline{$^{30}$University of Illinois at Chicago, Chicago,                
                   Illinois 60607}                                            
\centerline{$^{31}$Northern Illinois University, DeKalb, Illinois 60115}      
\centerline{$^{32}$Northwestern University, Evanston, Illinois 60208}         
\centerline{$^{33}$Indiana University, Bloomington, Indiana 47405}            
\centerline{$^{34}$University of Notre Dame, Notre Dame, Indiana 46556}       
\centerline{$^{35}$Purdue University, West Lafayette, Indiana 47907}          
\centerline{$^{36}$Iowa State University, Ames, Iowa 50011}                   
\centerline{$^{37}$University of Kansas, Lawrence, Kansas 66045}              
\centerline{$^{38}$Kansas State University, Manhattan, Kansas 66506}          
\centerline{$^{39}$Louisiana Tech University, Ruston, Louisiana 71272}        
\centerline{$^{40}$University of Maryland, College Park, Maryland 20742}      
\centerline{$^{41}$Boston University, Boston, Massachusetts 02215}            
\centerline{$^{42}$Northeastern University, Boston, Massachusetts 02115}      
\centerline{$^{43}$University of Michigan, Ann Arbor, Michigan 48109}         
\centerline{$^{44}$Michigan State University, East Lansing, Michigan 48824}   
\centerline{$^{45}$University of Nebraska, Lincoln, Nebraska 68588}           
\centerline{$^{46}$Columbia University, New York, New York 10027}             
\centerline{$^{47}$New York University, New York, New York 10003}             
\centerline{$^{48}$University of Rochester, Rochester, New York 14627}        
\centerline{$^{49}$State University of New York, Stony Brook,                 
                   New York 11794}                                            
\centerline{$^{50}$Brookhaven National Laboratory, Upton, New York 11973}     
\centerline{$^{51}$Langston University, Langston, Oklahoma 73050}             
\centerline{$^{52}$University of Oklahoma, Norman, Oklahoma 73019}            
\centerline{$^{53}$Brown University, Providence, Rhode Island 02912}          
\centerline{$^{54}$University of Texas, Arlington, Texas 76019}               
\centerline{$^{55}$Texas A\&M University, College Station, Texas 77843}       
\centerline{$^{56}$Rice University, Houston, Texas 77005}                     
\centerline{$^{57}$University of Washington, Seattle, Washington 98195}       
}                                                                             

\date{\today}
\maketitle
\pagebreak

\begin{abstract}
Using the D\O\ detector, 
we have studied events produced in \pbarp\ collisions that contain large 
forward regions with  very little energy deposition (``rapidity gaps'')
and concurrent jet production at center-of-mass energies of \eboth.
The fractions of forward and central jet events associated with such  
rapidity gaps are measured and compared to predictions from Monte Carlo 
models. 
For hard diffractive candidate events, we use the calorimeter to
extract the fractional momentum loss of the scattered protons.
  
\end{abstract}



Inelastic diffractive collisions are responsible for  10--15\% 
of the $\overline{p}p$ total 
cross section and have been described by Regge theory 
through the exchange of a pomeron~\cite{regge}. 
Diffractive events are characterized by the
absence of significant hadronic particle activity 
over a large region of rapidity or pseudorapidity ($\eta=-\ln[\tan(\frac{\theta}{2})]$, 
where $\theta$ is the polar angle relative to the beam).  
This empty region is 
called a rapidity gap and can be used as an experimental signature
for diffraction.  
Recent interest in diffraction has centered on 
the possible partonic nature of the pomeron in the framework 
of quantum chromodynamics (QCD), 
as suggested by Ingelman and Schlein~\cite{IS}.
Hard single diffraction (HSD),
which combines diffraction and a hard scatter
(such as jet or $W$-boson production), can be
used to study the properties of the pomeron.

The partonic nature of the pomeron 
was first inferred by the UA8 experiment~\cite{UA8} at the
CERN $Sp\overline{p}S$ collider at \cmelo\ from studies
of diffractive jet events.
Recent analyses of 
diffractive jet production~\cite{CDFJ,Zeus,H1} and
diffractive $W$-boson production~\cite{CDFW2}
are consistent with a predominantly hard gluonic pomeron,
but measured rates at the Fermilab Tevatron are several times lower
than predictions based on data from the DESY $ep$ collider HERA~\cite{fact}.
In this Letter we present new measurements of the characteristics
of diffractive jet events, and of the fraction 
of central and forward jet events that contain forward rapidity 
gaps (``gap fraction'') at
center-of-mass energies \eboth.
These measurements augment previous results from
the CDF collaboration on the gap fraction for 
forward jets at \cmehi~\cite{CDFJ}
and place further constraints on diffractive models.

In the D\O\ detector~\cite{NIM},
jets are measured using
the uranium/liquid-argon calorimeters with an electromagnetic section
extending to $|\eta|\!<\!4.1$ and coverage for 
hadrons to $|\eta|\!<\!5.2$. Jets are reconstructed using 
a fixed-cone algorithm with radius ${\cal R} = \sqrt{\Delta\eta^2 + \Delta\phi^2}=0.7$
($\phi$ is the azimuthal angle). The jets are 
corrected using  
standard D\O\ routines for jet-energy scale~\cite{D0jets},
except that there is
no subtraction of energy from spectator parton interactions, since
these are unlikely for diffractive events.

To identify rapidity gaps, we measure the number of tiles containing
a signal in the L\O\ forward scintillator arrays (\nl), and 
towers 
($\Delta\eta \XX \Delta\phi = 0.1 \XX 0.1$)
above threshold in the calorimeters (\ncal).
The L\O\ arrays provide  partial coverage in the region $2.3<|\eta|<4.3$.
A portion of the two forward calorimeters ($3.0<|\eta|<5.2$)
is used to measure the calorimeter multiplicity, with
a particle tagged by the deposition of more than 
$150$ (500)\,MeV of energy in an electromagnetic 
(hadronic) calorimeter tower. 
The thresholds are set to give
negligible noise from uranium decays, while maximizing 
sensitivity to energetic particles~\cite{Kristal}.

For $\rs=630$ and $1800$ GeV, we use triggers which required at least two 
jets with transverse energy 
$E_T > 12$ or $15$ GeV (see Table~\ref{events}) to study the
dependence of the gap fraction on jet location.
The forward jet triggers required the two leading jets to both have 
$\eta  > 1.6$ (or $\eta  < -1.6$), while the central jet triggers
had an offline requirement of $|\eta|<1.0$.
These data were obtained during
special low luminosity runs, with 
typical instantaneous luminosities much less than 
$1 \XX 10^{30}$ cm$^{-2}$s$^{-1}$.
At each \rs, we also implemented the so-called single veto trigger (SV),
a dijet trigger that required a rapidity gap on one side (using
the L\O\ detector). The SV trigger was used to obtain 
large samples of single diffractive candidate events. 
The events in the final data samples all have 
a single $\overline{p}p$ interaction requirement,
a vertex position within 50 cm of the center of the interaction region,
and two leading  jets that satisfy
standard quality criteria~\cite{incjet}.
The number of events in each of 
the final data samples 
and the integrated luminosities (${\cal L}$)
are given in Table~\ref{events}.

\begin{table} [htbp]
\begin{center}

\caption{Attributes of the final data samples.
}
\label{events}
\begin{tabular}{lcccr} 
Data Sample& Jet $|\eta|$  &Jet $E_T$(GeV) & ${\cal L}$ (nb$^{-1})$ & Events   \\   \hline
$1800$\,GeV Forward & $ >1.6$ & $>12$  & 62.9  & $50852$  \\ 
$1800$\,GeV Central & $ <1.0$ & $>15$  & 4.55  & $16567$  \\ 
$630$\,GeV  Forward & $ >1.6$ & $>12$  & 16.9  & $28421$   \\ 
$630$\,GeV  Central & $ <1.0$ & $>12$  & 8.06  & $48123$   \\ 
$1800$\,GeV SV   & $-$        & $>15$  & 5700  & $170393$  \\ 
$630$\,GeV  SV   & $-$        & $>12$  & 529   & $64772$   \\ 
\end{tabular}
\end{center}
\end{table}

\begin{figure}[ptbh]
\vspace{8mm}
\epsfxsize=4.0in
\centerline{\epsffile{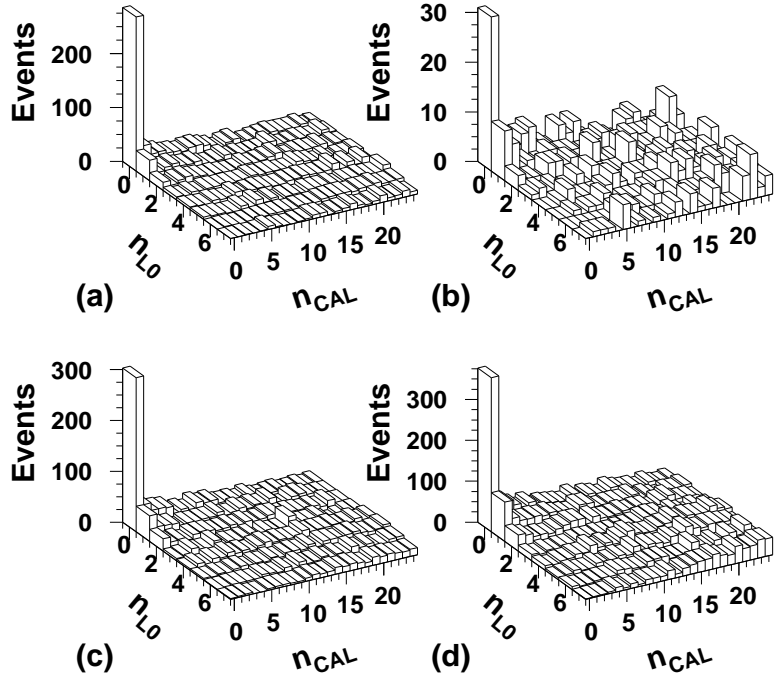}}
\vspace{12mm}
 \caption{Multiplicity distributions at \cmehi\ for (a) forward and (b) central
 jet events, and at \cmelo\ for (c) forward  and (d) central jet events.}
\label{raw_mult}
\end{figure}

The  \nl\
versus \ncal\ distributions for central and forward jet events at
$\sqrt{s}=630$ and $1800$\,GeV
are shown in Fig.~\ref{raw_mult}. For forward jet events,
these quantities are defined by the  $\eta$ region on the side opposite
the two leading jets, while for central jet events they are defined by the
forward $\eta$ interval that has the lower multiplicity.
The distributions display a peak at zero multiplicity 
($\ncal=\nl=0$),
in qualitative
agreement with expectations for a diffractive component in
the data.

\begin{figure}[hbtp]
\epsfxsize=4.0in
\centerline{\epsffile{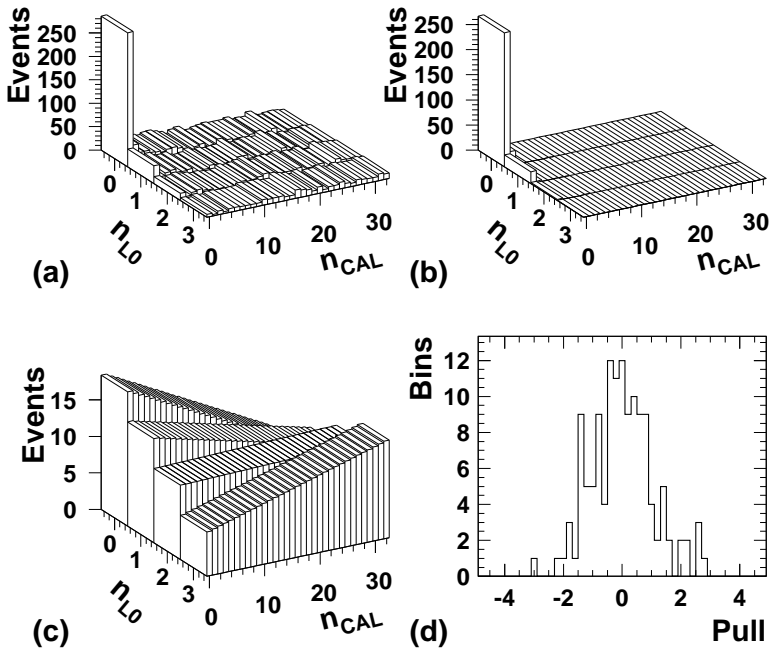}}
\vspace{4mm}
\caption{The (a) data from Fig.~\ref{raw_mult}(a), and corresponding
(b) fitted signal, (c) fitted background, 
and (d) normalized pull distributions.}
\label{fit_1800fwd}
\end{figure}

The gap fraction 
is extracted from a two-dimensional
fit to the lego plot of \nl\ versus \ncal.
The non-diffractive (high multiplicity) background is  
fitted in the signal region using a four-parameter polynomial, and 
the signal is fitted with a falling 
exponential, as suggested by Monte Carlo~\cite{Kristal}.  
Figure~\ref{fit_1800fwd} shows the multiplicity distribution
from Fig.~\ref{raw_mult}(a), and the resulting 
fitted signal, fitted background,
and normalized distribution of pulls ([data-fit]/$\sqrt{N}$).
All distributions have adequate fits, with  $\chi^2/{\rm dof}<1.2$.

Table~\ref{fitgapfrac} shows the gap fractions obtained for the 
four event samples.  The values range from $(0.22 \pm 0.05)\%$ 
for central jets
at $\rs=1800$\,GeV, to $(1.19 \pm 0.08)\%$ 
for forward jets at $\rs=630$\,GeV. 
Uncertainties are dominated by those on the fit parameters.
Additional small uncertainties from the dependence on the range of multiplicity
used in the fits were
added in quadrature. 
Potential sources of systematic error, such as
the number
of fit parameters, 
jet energy scale, trigger turn-on, tower threshold, 
luminosity, residual noise, and jet quality, yield only
negligible variations in the gap fractions~\cite{Kristal}.

Table~\ref{fitgapfrac} shows that 
the gap fractions at $\sqrt{s}=630$ GeV  are larger than  gap fractions at
\cmehi\
and that gap fractions for forward jets are larger than for central jets.
Table~\ref{fitgapfrac} also lists predicted gap fractions for several
possible pomeron structure functions
(discussed below).

We compare the data to Monte Carlo (MC) simulations using the 
hard diffractive event generator {\footnotesize POMPYT}~\cite{POMPYT},
which is based on the non-diffractive
{\footnotesize PYTHIA}~\cite{PYTHIA} program.
In {\footnotesize POMPYT},  a pomeron is emitted from the proton
with a certain probability (called the flux factor~\cite{IS}), and has
a structure functions $s(\beta)$,
where $\beta$ is the fractional momentum of the pomeron carried by the
hard parton.  
We used the standard Donnachie-Lanshoff
flux factor~\cite{DL} in this analysis
and compare our data to four structure 
functions: (i) ``hard gluon,'' a pomeron consisting of two gluons, 
$s(\beta)\propto \beta(1-\beta)$; (ii) ``flat gluon,'' 
$s(\beta)\propto \rm{constant}$; 
(iii) ``soft gluon,'' 
$s(\beta)
\propto (1-\beta)^5$;
and (iv) ``quark,''
the two-quark analog of (i).
In each case, the gap fraction is defined as the
cross section for jet events with a rapidity gap based on  
{\footnotesize POMPYT} divided by 
the jet cross section from {\footnotesize PYTHIA}
Many uncertainties, such as the
choice of proton parton densities, cancel in the ratio.
An MC version of the fitting method is applied to correct for
diffractive events that fail the gap selection criteria.
By applying the appropriate
correction factor (which ranges from a few per cent for soft gluon central jets
to about 80\% for hard gluon forward jets)
to each MC prediction and comparing 
to the data, we make no assumptions about which model (if any)
is correct~\cite{check}.

Monte Carlo gap fractions are 
shown in Table~\ref{fitgapfrac}. The systematic uncertainties
are 
typically dominated
by the difference in energy scale between data and Monte Carlo, 
but also include uncertainties from the fitting procedure.
We observe that rates for harder gluon structures 
are far higher than supported by data, 
while the quark structure is in reasonable agreement with the data.  
The quark structure, however, has previously been shown to predict an excessive
rate of diffractive $W$-Bosons~\cite{CDFW2}.

A hard gluonic pomeron is capable of describing previous 
measurements~\cite{CDFJ,Zeus,H1,CDFW2},
if combined with a
flux factor that decreases with increasing \rs~\cite{renorm}.
The ratios of gap fractions shown in the lower half of  Table~\ref{fitgapfrac}
provide new information, since 
the flux factor cancels for the same \rs, and
dependence on the flux factor is reduced for different \rs.
The  ratios for jets with $|\eta|>1.6$ to jets with 
$|\eta|<1.0$  show  clear disagreement
between the data and  
predictions for a hard-gluon pomeron structure, 
despite this cancellation.
A gluon-dominated pomeron containing both soft and hard components,
combined with a reduced flux factor, 
could describe all the data samples.
\begin{table*}[htbp]
\caption{The measured and predicted gap fractions and their ratios.}
\label{fitgapfrac} \begin{center} \begin{tabular}{lccccc} 
\multicolumn{6}{c}{~Gap Fractions} \\ 
Sample & Data & Hard Gluon & Flat Gluon & Soft Gluon & Quark \\
\hline
 $1800$\,GeV  $|\eta|>1.6$ & $(0.65 \pm  0.04)\% $  &
     $(2.2 \pm 0.3)\%$ & $(2.2 \pm 0.3)\%$ & 
     $(1.4 \pm 0.2)\%$ & $(0.79 \pm 0.12)\%$ \\
 $1800$\,GeV  $|\eta|<1.0$ & $(0.22 \pm 0.05\%$   &
     $(2.5 \pm 0.4)\%$ & $(3.5 \pm 0.5)\%$ &
     $(0.05 \pm 0.01)\%$ & $(0.49 \pm 0.06)\%$ \\
 $630$\,GeV  $|\eta|>1.6$  & $(1.19 \pm 0.08)\%$   &
     $(3.9  \pm 0.9)\%$ & $(3.1 \pm 0.8)\%$ &
     $(1.9  \pm 0.4)\%$ & $(2.2 \pm 0.5)\%$ \\ 
 $630$\,GeV  $|\eta|<1.0$  & $(0.90 \pm 0.06)\%$   &
     $(5.2 \pm 0.7)\%$ & $(6.3 \pm 0.9)\%$ &
     $(0.14 \pm 0.04)\%$ & $(1.6 \pm 0.2)\%$ \\ \hline
\multicolumn{6}{c}{~Ratios of Gap Fractions} \\ \hline
 $630/1800$ $|\eta|>1.6$  & $1.8 \pm 0.2$  &
     $1.7 \pm 0.4$ & $1.4 \pm 0.3$ & 
     $1.4 \pm 0.3$ & $2.7 \pm 0.6$ \\
 $630/1800$ $|\eta|<1.0$  & $4.1 \pm 0.9$   &
     $2.1 \pm 0.4$ & $1.8 \pm 0.3$ &
     $3.1 \pm 1.1$ & $3.2 \pm 0.5$ \\
 $1800$ $|\eta|>1.6/|\eta|<1.0$ & $3.0 \pm 0.7 $   &
     $0.88 \pm 0.18$ & $0.64 \pm 0.12$ &
     $30. \pm 8.$ & $1.6 \pm 0.3$ \\ 
 $630$ $|\eta|>1.6/|\eta|<1.0$   & $1.3 \pm 0.1$   &
     $0.75 \pm 0.16$ & $0.48 \pm 0.12$ &
     $13. \pm 4.$ & $1.4 \pm 0.3$ \\       
  \end{tabular}
 \end{center}
\end{table*}

The characteristics of the HSD events were examined
using the high statistics 
SV trigger.  
\begin{figure}[ht]
\epsfxsize=4.0in
\centerline{\epsffile{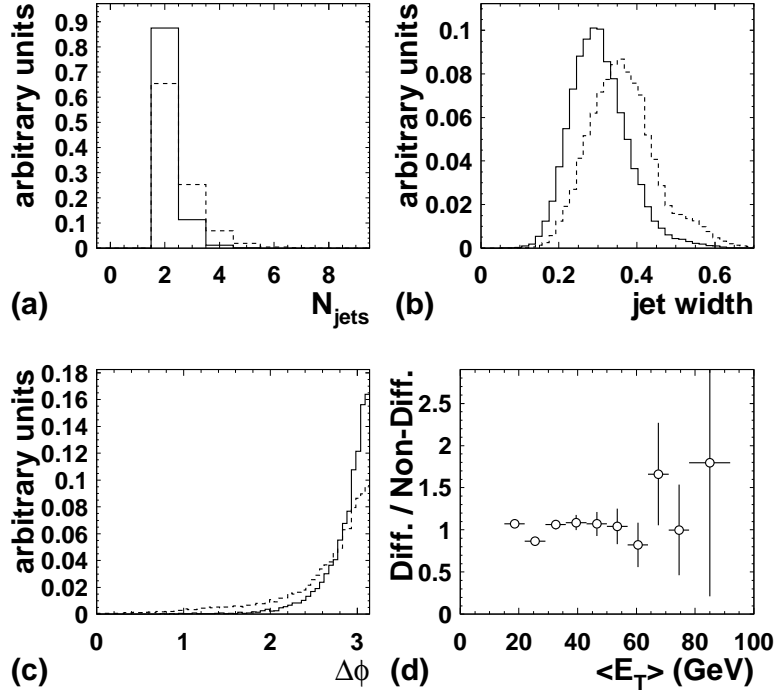}}
\vspace{4mm}
 \caption{Distributions of the (a) number of jets, 
(b) jet width, (c)  $\Delta \phi$
between leading jets, for central 
diffractive (solid) and non-diffractive (dashed) jet events at \cmehi.
(d) The relative ratio of diffractive to non-diffractive
events as a function of the average
 $E_T$ of the two leading jets.}
\label{char}
\end{figure}
We plot in Fig.~\ref{char} the distributions of the  number of jets, the
$E_T$-weighted rms jet widths, the $\Delta \phi$ between the two leading jets, 
and the relative 
ratio of diffractive to non-diffractive 
events as a function of the average $E_T$ of the two leading jets,
for central jets at \cmehi. The 
solid lines in Fig.~\ref{char}(a)--(c) correspond to the distributions 
for HSD candidate events  ($\ncal = \nl = 0$),
and the dashed lines show the distributions for 
non-diffractive events ($\ncal > 0$ and $\nl > 0$).  These
plots show that 
the diffractive events appear to have less overall radiation.  
Figure~\ref{char}(d) indicates that there is
little dependence of the gap fraction on  average jet $E_T$.
The MC samples (not shown) have characteristics similar to the data.

Finally, we measure the fractional momentum loss of 
the proton $\xi$, defined as~\cite{collins}:

\begin{equation}
\xi \approx \frac{1}{\sqrt{s}}\sum_{i} E_{T_i}e^{\eta_i} \\
\end{equation}

\noindent where the summation is over all observed
particles.
The outgoing scattered proton or antiproton 
(and the rapidity gap) is defined to be 
at positive $\eta$. Equation~(1) weights heavily the well-measured central
region near the rapidity gap, while particles that escape down
the beam pipe at negative $\eta$ 
give a negligible contribution.
Using  {\footnotesize POMPYT} events,
where $\xi$ can be determined
from the momentum of the scattered proton, 
we have verified that Eq.~(1) is reliable 
at both values of \rs\ and for different 
pomeron structures. A scale factor ($2.2 \pm 0.3$) derived from
Monte Carlo is used to convert
$\xi$ measured from all  particles to that from
just electromagnetic calorimetric energy depositions~\cite{Kristal}. 
\begin{figure}[ht]
\epsfxsize=4.0in
\centerline{\epsffile{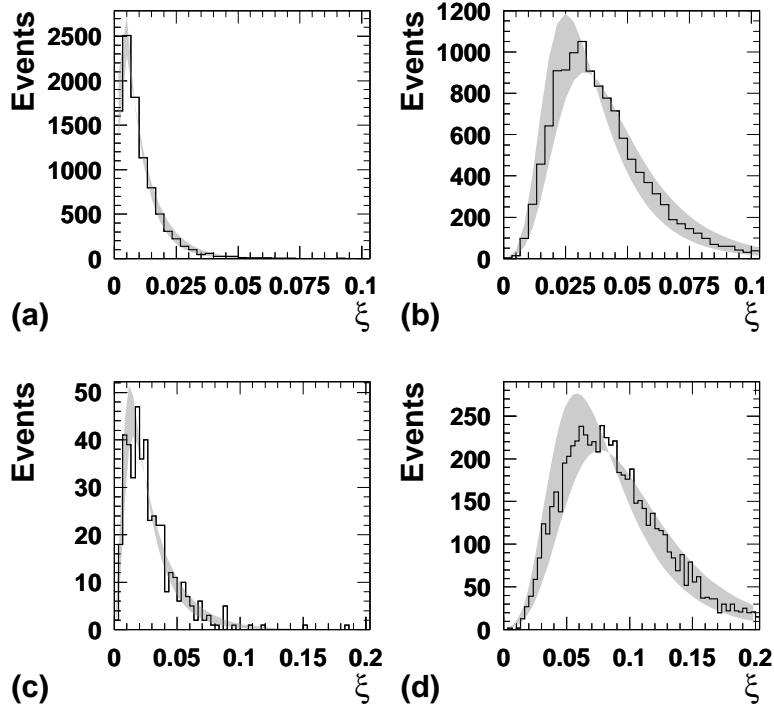}}
\vspace{4mm}
\caption{The $\xi$ distributions  for
$\rs=1800$\,GeV (a) forward and (b) central
jets and for  $\rs=630$\,GeV (c) forward and (d) central jets,
using the SV trigger with $\ncal = \nl = 0$.
The shaded region shows the variance in the distribution due
to energy scale uncertainties (see text).}
\label{svxi}
\end{figure}
The $\xi$ distributions for
 forward and central jets at
\eboth\ are displayed 
in Fig.~\ref{svxi}, with the shaded region
showing the variance in the distribution due to energy scale uncertainties.
Energy-scale uncertainties result in a shift in $\xi$ such that if the true
distribution were below the histogram at small $\xi$, it would be above
the histogram at large $\xi$.

The $\xi$ distributions show the expected kinematic behavior
of diffraction ($M=\sqrt{\xi s}$, where $M$ is the mass of the
diffractive system), peaking
at larger $\xi$ for central jets than for forward jets.
Forward and central jets
at \cmelo\ also peak at larger $\xi$ values with respect to the
corresponding distributions at \cmehi, since for fixed diffractive
mass, smaller \rs\ implies larger $\xi$.
Even though pomeron exchange is thought to dominate only for  $\xi < 0.05$,
the trends of the $\xi$ distributions 
can be reproduced by {\footnotesize POMPYT}.
Without the observation of the scattered proton, the interpretation
of these large 
$\xi$ rapidity gap events is uncertain.

We have measured  properties of hard single diffraction 
at \eboth\ with jets at forward and central rapidities.
The gap fractions have been measured without applying model-dependent
corrections. Within the Ingelman-Schlein model, our data 
can be reasonably described by 
a pomeron composed dominantly of quarks. 
For the model to describe our data as well as previous measurements,
a reduced flux factor convoluted with 
a gluonic pomeron containing significant soft and hard components
is required. 
We have also measured the fractional momentum 
lost by the scattered proton and found it greater than typically 
expected for pomeron exchange.

%
We thank the Fermilab and collaborating institution staffs for 
contributions to this work, and acknowledge support from the 
Department of Energy and National Science Foundation (USA),  
Commissariat  \` a L'Energie Atomique (France), 
Ministry for Science and Technology and Ministry for Atomic 
   Energy (Russia),
CAPES and CNPq (Brazil),
Departments of Atomic Energy and Science and Education (India),
Colciencias (Colombia),
CONACyT (Mexico),
Ministry of Education and KOSEF (Korea),
CONICET and UBACyT (Argentina),
A.P. Sloan Foundation,
and the Humboldt Foundation.


\begin{thebibliography}{99}
\bibitem{check}  
Ref.~\cite{CDFJ} obtained a gap fraction 
of $(0.75 \pm 0.10)\%$
assuming a hard gluon structure. 
If we correct our corresponding 
measurement of $(0.65 \pm 0.04)\%$ in a comparable manner, we obtain a
consistent value of $(0.88 \pm 0.05)\%$.
\end{thebibliography}
\end{document}